\documentclass[aps,a4paper,twocolumn,prl,floatfix,superscriptaddress]{revtex4-1}
\usepackage{listings}
\usepackage[T1]{fontenc}
\usepackage[utf8]{inputenc}
\usepackage{newtxtext}
\usepackage{amsmath}
\usepackage{mathtools}
\usepackage{amsfonts}
\usepackage{bm}
\usepackage{amssymb}

\usepackage{newtxmath}
\usepackage{color}
\usepackage{xcolor}
\definecolor{blue}{rgb}{0.2, 0.3, 0.85}
\definecolor{red}{RGB}{0,100,0}
\definecolor{darkgreen}{rgb}{0.0, 0.5, 0.0}
\usepackage[colorlinks,bookmarks=false,citecolor=red,linkcolor=blue,urlcolor=blue]{hyperref}
\renewcommand{\section}[1]{\noindent{\textit{\textbf{#1---}}}}
\renewcommand{\subsection}[1]{{\textit{#1.~}}}
\usepackage{graphicx} %Loading the package

\usepackage{tikz}
\usepackage{textcomp}

\def\be{\begin{equation}}
\def\ee{\end{equation}}
\def\bea{\begin{eqnarray}}
\def\eea{\end{eqnarray}}

\def\ii{\rm{i}}

\begin{document}
\title{Resilient Infinite Randomness Criticality for a Disordered Chain of Interacting  Majorana Fermions}
\author{Natalia Chepiga}
\affiliation{Kavli Institute of Nanoscience, Delft University of Technology, Lorentzweg 1, 2628 CJ Delft, The Netherlands}
\author{Nicolas Laflorencie}
\affiliation{Laboratoire de Physique Th\'{e}orique, Universit\'{e} de Toulouse, CNRS, UPS, France}
%\date{\today}
\begin{abstract}
The quantum critical properties of interacting  fermions in the presence of disorder are still not fully understood.  While it is well known that for Dirac fermions, interactions are irrelevant to the non-interacting infinite randomness fixed point (IRFP), the problem remains largely open in the case of Majorana fermions which further display a much richer disorder-free phase diagram. Here, pushing the limits of DMRG simulations, we  carefully examine the ground-state of a Majorana chain with both disorder and interactions. Building on  appropriate boundary conditions and  key observables such as entanglement, energy gap,  and correlations, we strikingly find that the non-interacting Majorana IRFP is very stable against finite interactions, in contrast with previous claims.
\end{abstract}
\maketitle

\section{Introduction} The interplay of disorder and interactions in low dimensional systems is one of the most fascinating problem of condensed matter physics, with highly non-trivial open questions, the many-body localization (MBL) being a remarkable example~\cite{alet_many-body_2018,abanin_many-body_2019}. One of the key points of MBL physics concerns the stability of a non-interacting Anderson insulator against interactions at (in)finite temperature, a question already raised in the pioneering works~\cite{fleishman_interactions_1980,gornyi_interacting_2005,basko_metalinsulator_2006}. 
Since then, a significant and flourishing activity has continued to explore these questions, but with controversial predictions~\cite{sierant_thouless_2020,abanin_distinguishing_2021,brighi_propagation_2022,sierant_challenges_2022,sels_bath-induced_2022,morningstar_avalanches_2022}. 

In this work, we propose to take a small detour by focusing on the different but closely related problem of the low-energy properties of the interacting Majorana chain (IMC) model~\cite{rahmani_emergent_2015,rahmani_phase_2015,milsted_statistical_2015,karcher_disorder_2019,chepiga_topological_2022} in the presence of disorder. It is governed by the following one-dimensional (1D) Hamiltonian
\be
{\cal{H}}=-\sum_j\left({\rm{i}}t_j\gamma_{j}\gamma_{j+1}+g\gamma_{j}\gamma_{j+1}\gamma_{j+2}\gamma_{j+3}\right),
\label{eq:IMC}
\ee
with random couplings $t_j$ and constant interaction $g$. The operators $\gamma_{j}$ are Majorana (real) fermions ($\gamma_{j}^{\vphantom{\dagger}}=\gamma_{j}^{\dagger}$ and $\{\gamma_{i},\gamma_{j}\}=2\delta_{ij}$) from which Dirac (complex) fermions can be constructed as pairs of Majoranas such that $2c_j=\gamma_{2j-1}+{\rm{i}}\gamma_{2j}$, yielding the Dirac fermions version of the IMC model Eq.~\eqref{eq:IMC} which can also be seen as the interacting counterpart of the Kitaev chain model~\cite{kitaev_unpaired_2001,sm}. There is a third possible formulation in terms of Pauli matrices~\cite{sm} 
\be
{\cal{H}}=\sum_\ell\left[J_{\ell}\sigma_{\ell}^{x}\sigma_{\ell+1}^{x}+h_{\ell}\sigma^{z}_{\ell}+g\left(\sigma_{\ell}^{z}\sigma_{\ell+1}^{z}+\sigma_{\ell}^{x}\sigma_{\ell+2}^{x}\right)\right],
\label{eq:IMCI}
\ee
with $J_{\ell}=t_{2j}$ and $h_{\ell}=t_{2j-1}$. In the absence of interactions ($g=0$), this problem simply boils down to the celebrated transverse field Ising chain (TFI) model~\cite{pfeuty_one-dimensional_1970}. In the random case, if couplings and fields are such that ${\overline{\ln J}}={\overline{\ln h}}$ (where ${\overline{[\cdots]}}$ stands for disorder averaging), the so-called infinite-randomness fixed point (IRFP)~\cite{fisher_random_1992,fisher_critical_1995,igloi_strong_2005} describes the physics, as carefully checked numerically both for ground-state~\cite{young_numerical_1996,henelius_numerical_1998} and excited states~\cite{huang_excited-state_2014,laflorencie_entanglement_2022}.
\vskip 0.15cm
\section{Infinite-randomness hallmarks} 
To fix the context, we first list some key properties of the 1D IRFP. (i) Time and space are related in a strongly anisotropic way, with a dynamical critical exponent $z=\infty$. As a result the lowest energy gap $\Delta$ does not self-average, is broadly distributed, and exponentially suppressed with the chain length $N$, such that 
\be
{\overline{\ln \Delta}}\sim -\sqrt{N}.
\label{eq:gap}
\ee
(ii) There is also lack of self-averaging for the spin-spin correlations: the average decays algebraically, while the typical vanishes much faster, as a stretched exponential
\be
{\overline{\langle \sigma_\ell^x\sigma_{\ell+r}^{x}\rangle}}\sim {r^{\left({\sqrt{5}-3}\right)/{2}}}\quad {\rm{and}}\quad{\overline{\ln \,\langle \sigma_\ell^x\sigma_{\ell+r}^{x}\rangle}}\sim -\sqrt{r}.
\label{eq:corr}
\ee
(iii) Despite the absence of conformal invariance, the R\'enyi entanglement entropy (EE) grows logarithmically with the sub-system length $n$, as in the clean case~\cite{holzhey_geometric_1994,vidal_entanglement_2003,calabrese_entanglement_2004}, following  
\be
{\overline{S_{q}(n)}}=\frac{c_{\rm eff}}{6}\ln (n) + s_q,
\label{eq:EE}
\ee
for open boundaries, $s_q$ being a non-universal constant. The key object here is the so-called "effective central charge" $c_{\rm eff}$, which for the IRFP  is given by $c_{\rm eff}^{\rm IRFP}=c\ln 2$~\cite{refael_entanglement_2004,refael_criticality_2009,laflorencie_scaling_2005,hoyos_correlation_2007,fagotti_entanglement_2011}, where $c$ is the central charge of the underlying clean fixed point. 

%%%%%%%
\begin{figure*}
    \centering
    \includegraphics[width=2\columnwidth]{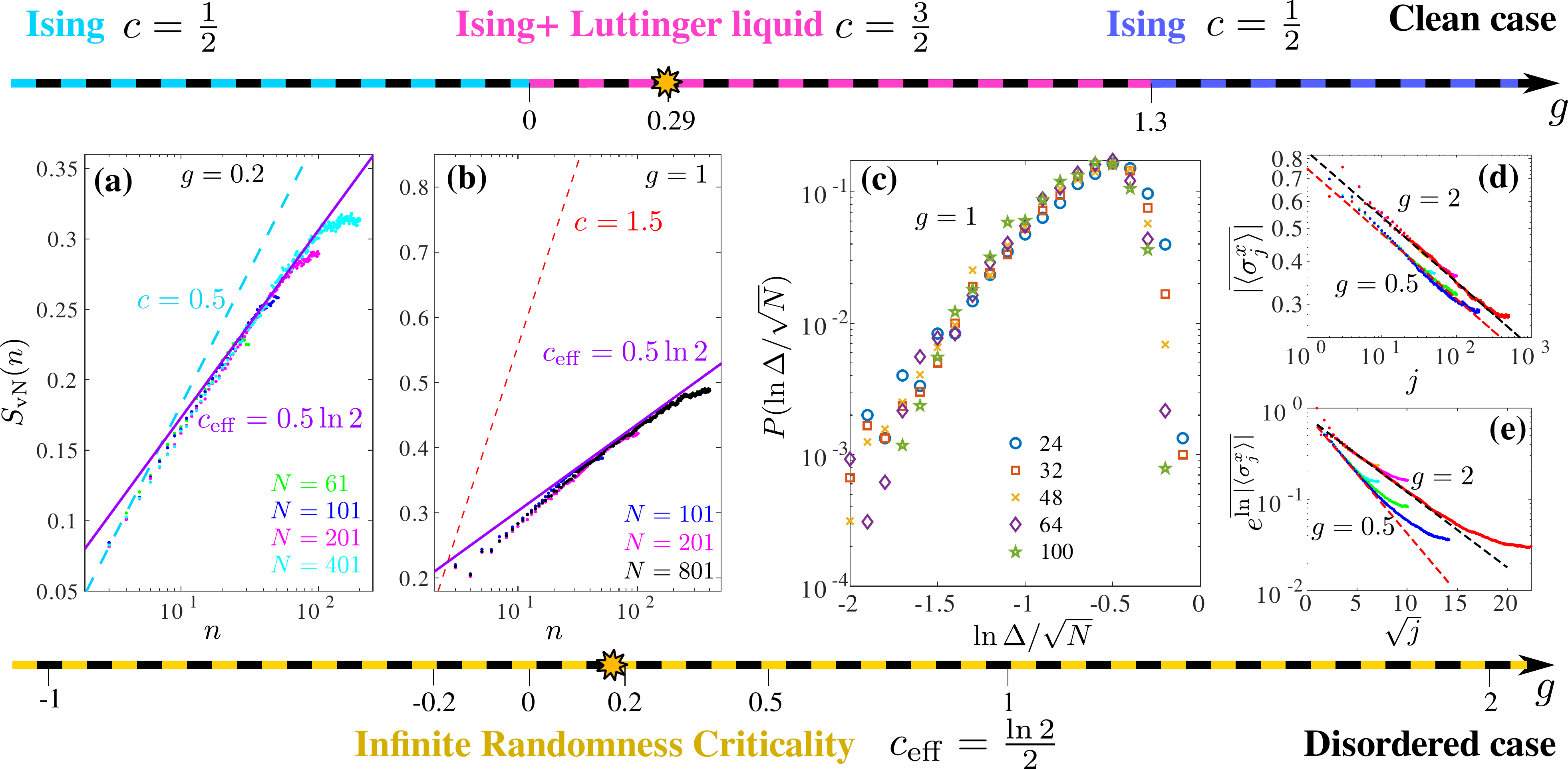}
        \caption{Overview of  the interacting Majorana chain model Eq.~\eqref{eq:IMC}. Top and bottom arrows present  the phase diagrams for both clean and disordered models. The clean case (see Ref.~\cite{chepiga_topological_2022}) displays three critical phases with central charges $c=1/2$ and $3/2$. Instead, the random case displays a unique Infinite Randomness Criticality, as demonstrated by representative cases in the various panels. {\bf{(a-b)}} show the von-Neumann entanglement entropy $S_{\rm {vN}}(n)$ scaling as a function of subsystem length $n$, for $g=0.2$ and $g=1$ for which the clean scalings (with $c=0.5$ and $c=1.5$) are compared with the disorder-average EE for various lengths $N$, which exhibit the IRFP scaling with $c_{\rm eff}=0.5\,{\ln 2}$ (see also Fig.~\ref{fig:ee} below). Panel {\bf{(c)}} presents another smoking gun of IRFP with the universal collapse for the distribution of the lowest gap $P\left(\frac{\ln \Delta}{\sqrt{N}}\right)$, displayed  for $g=1$ and various system sizes $N$, see also Fig.~\ref{fig:gap}. Panels {\bf{(d-e)}} show the decay of the average and typical magnetizations, away form the boundary, for two representative cases $g=0.5$ and $g=2$ showing perfect agreement with IRFP criticality, see also Fig.~\ref{fig:corr} for more details and results. The yellow stars on top and bottom arrows denote the onset of incommensurability, further discussed in Fig.~\ref{fig:IC}.} \label{fig:overview}
\end{figure*}
%%%%%%%

Such an unbounded entanglement growth Eq.~\eqref{eq:EE} strongly contrasts with MBL or Anderson insulators for which a strict area law is observed, even at infinite temperature, with an EE bounded by the finite localization length~\cite{bauer_area_2013,laflorencie_entanglement_2022}. Here, the IRFP is only marginally localized, i.e., that all single-particle states have a finite localization length, except in the band center where the localization is stretched exponential~\cite{eggarter_singular_1978,fisher_random_1994,nandkishore_marginal_2014}.
\vskip 0.15cm
\section{IRFP and interactions} Two historical examples of non-interacting IRFPs are the 1D disordered TFI model~\cite{fisher_random_1992,fisher_critical_1995}, and the random-bond XX chain~\cite{fisher_random_1994}. Interestingly, both models can be seen as the opposite sides of the same coin: non-interacting Majorana (real) {\it{vs.}}  Dirac (complex) fermions with random hoppings. Although the effect of interactions was quickly understood as {\it{irrelevant}} in a Renormalization Group (RG) sense~\cite{doty_effects_1992,fisher_random_1994} for free Dirac fermions, the story turned out to be quite different in the case of Majoranas. In his seminal work, Fisher first suggested that interactions should also be {\it{irrelevant}} at the IRFP in the Ising/Majorana  case~\cite{fisher_critical_1995}, but this issue remained essentially unexplored for many years, before re-emerging only recently in the MBL context~\cite{pekker_hilbert-glass_2014,slagle_disordered_2016,you_entanglement_2016,monthus_strong_2018,sahay_emergent_2021,moudgalya_perturbative_2020,laflorencie_topological_2022,wahl_local_2022}. There at high energy, the IRFP was found to be destablized by weak interactions towards a delocalized ergodic phase~\cite{sahay_emergent_2021,moudgalya_perturbative_2020,laflorencie_topological_2022}. 

Despite these progress made at high energy,  the status of the ground-state of the disordered IMC model Eq.~\eqref{eq:IMC} is still controversial, with rather intriguing recent conclusions~\cite{milsted_statistical_2015,karcher_disorder_2019} contrasting with previous  claims~\cite{fisher_critical_1995}. Building on DMRG simulations Milsted {\it et al.}~\cite{milsted_statistical_2015} observed a saturation of the EE for repulsive interaction $g>0$, in agreement with Karcher {\it et al.}~\cite{karcher_disorder_2019} who further concluded that the system gets localized and spontaneously breaks the duality symmetry of the IMC Hamiltonian, for any $g>0$. Results in the attractive regime $g<0$, again based on EE scaling, are more ambiguous: Ref.~\cite{milsted_statistical_2015} concludes that IRFP is stable, while Ref.~\cite{karcher_disorder_2019} states on the contrary that disorder becomes irrelevant and that the clean fixed point physics is recovered. 
\vskip 0.15cm
\section{Main results and phase diagram} Our work falls within this puzzling and stimulating context. By pushing the limits of DMRG simulations for disordered quantum systems~\cite{note_dmrg}, we carefully and deeply explore the ground-state properties of the IMC model Eq.~\eqref{eq:IMC} in the presence of both interactions and  randomness. Our main result, summarized in Fig.~\ref{fig:overview}, is that the IRFP is robust and stable to finite interactions. While in the clean case~\cite{rahmani_phase_2015,chepiga_topological_2022}, a succession of critical phases is observed upon varying $g$, with  central charges $c=1/2,\,3/2$, adding disorder to the Majorana hopping terms is a {\it{relevant}} perturbation. For the range of interactions considered in this work, the non-interacting IRFP appears to be the unique attractive fixed point, thus reinforcing the original expectation~\cite{fisher_critical_1995} that interactions are therefore {\it{irrelevant}} to the free Majorana IRFP.

Our conclusions are based on the complementarity of key observables used to probe the various aforementioned properties of the IRFP. This is exemplified in Fig.~\ref{fig:overview} where the von-Neumann EE (a-b), the low-energy gap (c), and the average and typical order parameters  (d-e) are displayed across the various regimes of interaction strength, all panels showing one of the smoking gun feature characteristic of the IRFP.

 In the rest of the work, we present and discuss very carefully our numerical results building on these three pivotal observables, several technical aspects being detailed in the supplementary material~\cite{sm}. Let us however mention that we simulate the IMC model Eq.~\eqref{eq:IMC} in its "magnetic" version Eq.~\eqref{eq:IMCI}, and mostly focus on the repulsive $g>0$ regime. Although interesting effects are certainly expected away from it, we stick to the self-dual line  ${\overline{\ln J}}={\overline{\ln h}}$, independently drawing  $J_i$ and $h_i$ from a box $[1-W,1+W]$ with $W=0.9$~\cite{noteW}. A very important issue, sometimes  overlooked, concerns the number of random samples which we take as large as possible (typically between $3000$ and $8000$). This is particularly meaningful at IRFPs where rare events play a pivotal role, and broad distributions are crucially important  to describe the physics.

%%%%%%%
\begin{figure}[bp]
    \centering
    \includegraphics[width=\columnwidth,clip]{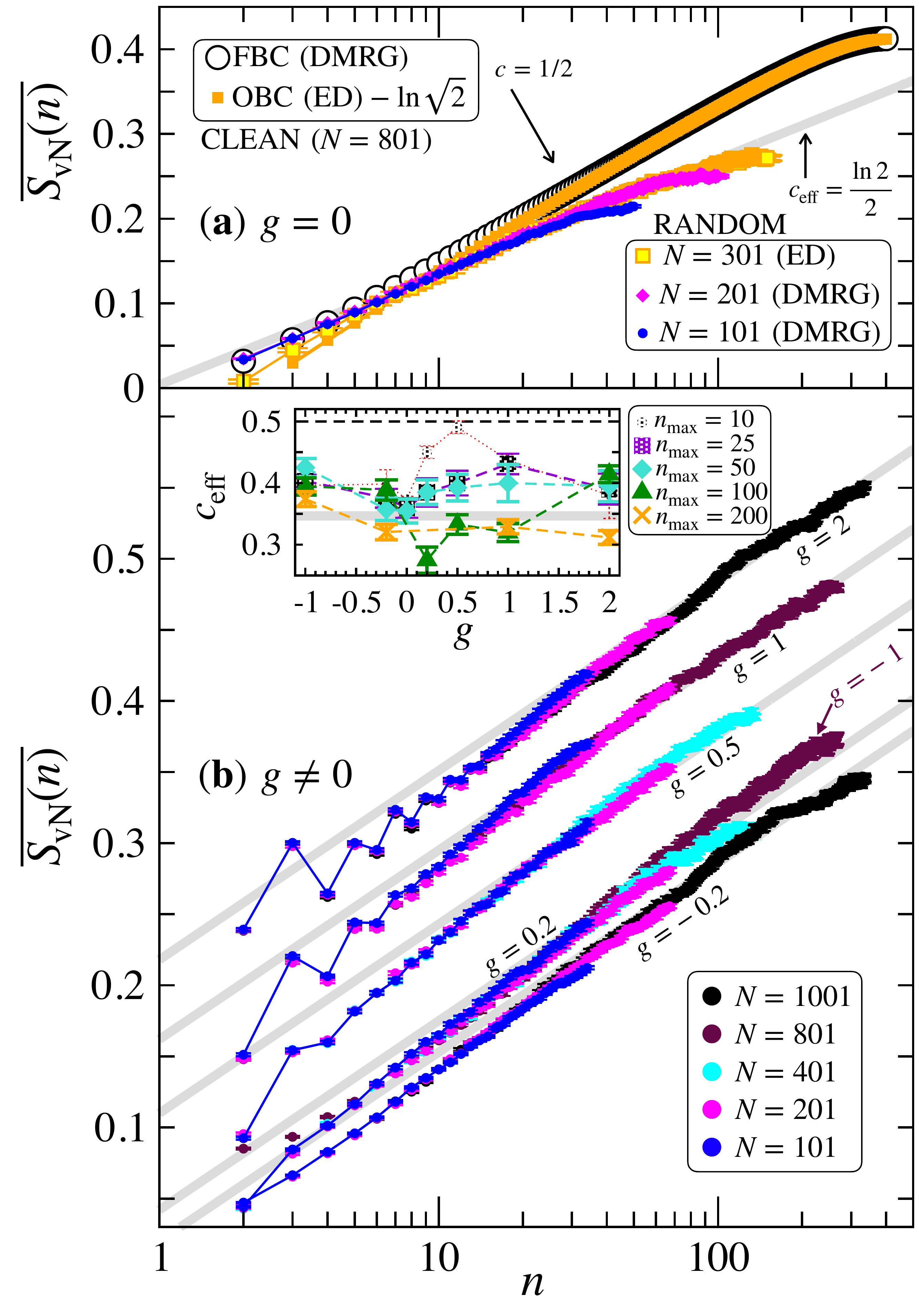}
        \caption{DMRG and ED results for the von-Neumann entropy scaling as a function of sub-system size $n$ for (a) non-interacting, and (b) interacting Majorana fermions, Eq.~\eqref{eq:IMC}. {\bf (a)} $g=0$, clean chain results (upper data)  illustrate how OBC ED data match with FBC DMRG (after subtracting the boundary entropy $\ln\sqrt{2}$). In the random case, a similar agreement is observed for the disorder-average (after the same subtraction), the dominant scaling being now controlled by Eq.~\eqref{eq:EE} with an "effective central charge" $c_{\rm eff}=\frac{\ln 2}{2}$ (grey line), a finite-size bending down is observed when half-chain is approached. {\bf (b)} $g\neq 0$ DMRG results  shown for subsystems $2\le n\le N/3$, various interaction strengths (indicated on the plot), and different chain lengths (colored symbols). The agreement with the IFRP scaling (grey line Eq.~\eqref{eq:EE} with $c_{\rm eff}=\frac{\ln 2}{2}$) is excellent in all cases, once the asymptotic regime is reached beyond a finite crossover length scale~\cite{laflorencie_crossover_2004,laflorencie_entanglement_2022}. Inset: $g$-dependence of $c_{\rm eff}$ extracted from fits to the form Eq.~\eqref{eq:EE} over successive sliding windows ending at $n_{\rm max}$. All data agree with the asymptotic log scaling controlled by the prefactor $c_{\rm eff}=\frac{\ln 2}{2}$.} 
        \label{fig:ee}
\end{figure}
%%%%%%%

\vskip 0.15cm
\section{Entanglement entropy} 
Before getting to the EE itself, we start with a brief discussion of the boundary conditions, illustrated for the non-interacting case in Fig.~\ref{fig:ee} (a). Instead of open boundary conditions (OBC), most commonly used in the DMRG realm, here we shall use the so-called fixed boundary conditions (FBC), obtained by locally pinning the boundary spins with a strong longitudinal field~\cite{zhou_entanglement_2006,affleck_entanglement_2009}, thus artificially breaking the parity symmetry of the IMC Hamiltonian. As a result,  the FBC entropy is reduced from its OBC value by the Affleck-Ludwig boundary  term~\cite{affleck_universal_1991}, such that 
$S_{\rm vN}^{\rm FBC}=S_{\rm vN}^{\rm OBC}-\ln \sqrt{2}$, but does not loose its universal logarithmic scaling. This becomes clear in Fig.~\ref{fig:ee} (a) for free fermions ($g=0$) where DMRG and exact diagonalization (ED) data are successfully compared in the clean case.  Interestingly, we further observe that such a boundary entropy also shows up for the free-fermion IRFP, as evidenced in the same panel (a) of Fig.~\ref{fig:ee} where OBC ED data match with FBC DMRG after a subtraction of  the similar $\ln \sqrt{2}$ term.

Let us now present the most important result of the paper, displayed in Fig.~\ref{fig:ee} (b) where for finite interaction strengths $g\neq 0$, the disorder-average EEs show excellent agreement with the non-interacting IFRP logarithmic growth Eq.~\eqref{eq:EE}, with $c_{\rm eff}=\frac{\ln 2}{2}$. Remarkably, this remains true for the entire  regime of study $-1\le g\le 2$. This is even more clear from the inset where  the $g$-dependence of $c_{\rm eff}$ is extracted from fits to the form Eq.~\eqref{eq:EE} over successive sliding windows. This result deeply contrasts with previous works~\cite{milsted_statistical_2015,karcher_disorder_2019} where a saturation of EE was observed and interpreted as a consequence of localization. There are two main causes for this disagreement, both due to numerical limitations that most probably led to a misinterpretation of earlier DMRG data. The first reason is the number of kept DMRG states, which can be a major obstacle~\cite{note_dmrg}. The second, perhaps more interesting, comes from  the boundary conditions and our choice of FBC, which leads to a significant reduction in EE, giving a decisive advantage to  our DMRG simulations~\cite{sm}.

It is furthermore noteworthy that all finite interaction results show the same tendency to flow to the non-interacting IRFP scaling, with a unique effective central charge fully compatible with $c_{\rm eff}=\frac{\ln 2}{2}$, even in the repulsive regime where the clean case displays $c=3/2$ for $0.29\le g\le 1.3$, as clearly visible in Fig.~\ref{fig:overview} (b) for a comparison between clean and disordered cases at $g=1$.

\vskip 0.15cm
\section{Low-energy gap}  In order to double-check the IRFP hypothesis over the broad regime of interaction strengths, we also focus on the lowest energy gap $\Delta$ above the ground-state, and in particular we aim to check the very peculiar exponentially activated scaling law defined by Eq.~\eqref{eq:gap}, which signals a dynamical exponent $z=\infty$. In addition, the probability distribution of these gaps is expected to display broadening and a universal scaling form, as shown for free fermions~\cite{young_numerical_1996,fisher_distributions_1998}. 

\begin{figure}[tp]
    \centering
    \includegraphics[width=.78\columnwidth]{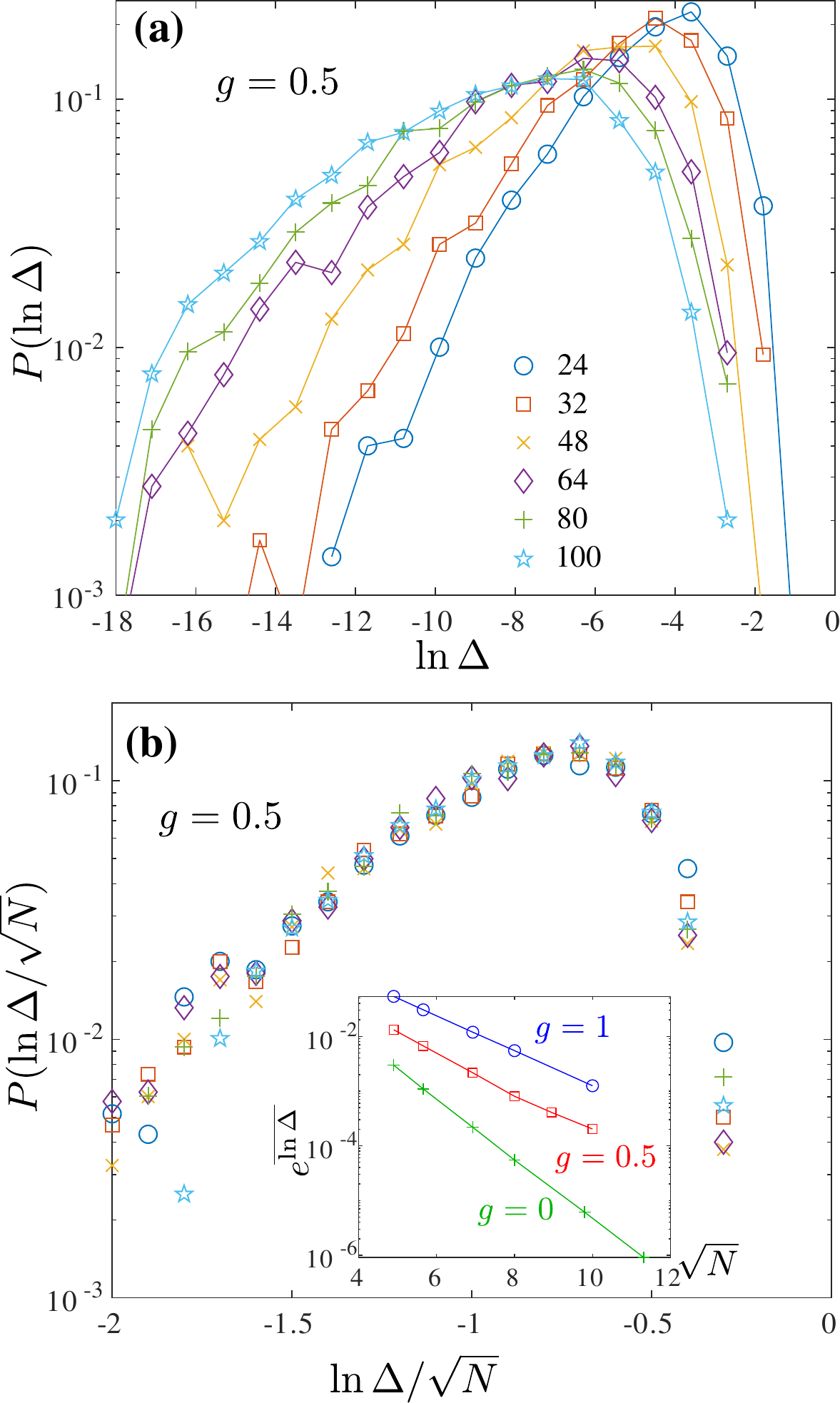}
        \caption{DMRG results for the lowest energy gap $\Delta$. {\bf{(a)}} Distribution  $P(\ln \Delta)$  collected at $g=0.5$ from $4000$ samples for various system sizes, as indicated on the plot. The broadening upon increasing $N$ is an IRFP signature, best seen in {\bf{(b)}} where the distributions of rescaled gaps $P\left({\ln \Delta}/{\sqrt{N}}\right)$ show very good collapse. Inset: the typical gap plotted {\it{vs.}} $\sqrt{N}$ for $g=0,\,0.5,\,1$, shows perfect agreement with the activated IRFP scaling Eq.~\eqref{eq:gap}.} \label{fig:gap}
\end{figure}

Here for the interacting model, we also observe, see Fig.~\ref{fig:gap} (a) for $g=0.5$, a very clear broadening of the distributions $P(\ln \Delta)$ upon increasing the system size, which is a strong evidence that $z=\infty$, as predicted for the IRFP. Furthermore, the same data show an excellent collapse in Fig.~\ref{fig:gap} (b) when histogrammized against $({\ln \Delta})/{\sqrt{N}}$, without  any adjustable parameter. We have checked that this remains true for other values of the interaction strength (in the range of study), as shown for a few values of $g$ in the inset of Fig.~\ref{fig:gap} (b). There, one sees that the typical gap $e^{\overline{\ln\Delta}}$ perfectly obeys the activated scaling law Eq.~\eqref{eq:gap}. The non-interacting case (ED data for $g=0$) is also displayed for comparison.

%%%%%%%
%%%%%%%

\vskip 0.15cm
\section{Correlations} 
\begin{figure}[tp]
    \centering
    \includegraphics[width=.95\columnwidth,clip]{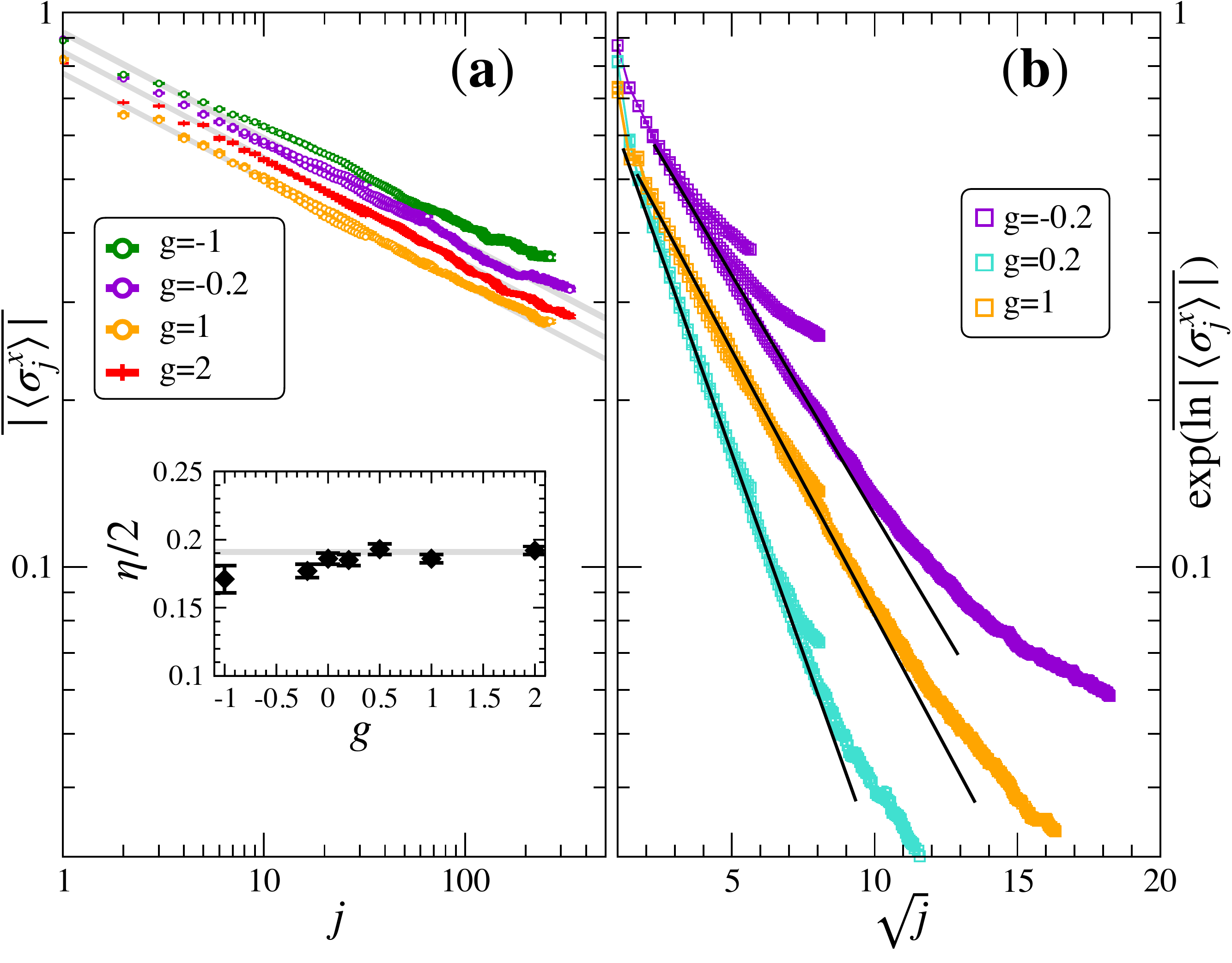}
        \caption{DMRG results for the decrease of the order parameter away from the boundary $j$, see Eq.~\eqref{eq:corrx}. {\bf{(a)}} Power-law decay of the disorder-average ${\overline{|\langle \sigma_j^x\rangle|}}$ shown for 4 different values of the interaction $g$, all in excellent agreement with the IRFP prediction $\eta/2=1-\phi/2\approx 0.191$ (grey line). Inset: estimated exponent $\eta/2$ plotted against $g$. {\bf{(b)}} The stretched exponential vanishing of the typical value $\exp({{\overline{\ln \,|\langle \sigma_j^x\rangle|}}})$ is fully compatible with the IRFP scaling (black lines).}\label{fig:corr}
\end{figure}

The last evidence for infinite randomness physics is captured by the spin correlations, as given by Eq.~\eqref{eq:corr}. The absence of self-averaging  is again reflected here in the clear qualitative difference between mean and typical decays of pairwise correlations:  power-law with a universal exponent $\eta={\frac{3-\sqrt{5}}{2}}\approx 0.382$ {\it{vs.}} stretched exponential. 
This IRFP feature can also be nicely captured with FBC. Indeed, when the edge spins are fixed, the following decrease of the order parameter is expected away from the boundary
\be
{\overline{|\langle \sigma_j^x\rangle|}}\sim {j^{-\eta/2}}
\quad {\rm{and}}\quad{\overline{\ln \,|\langle \sigma_j^x\rangle|}}\sim -\sqrt{j}.
\label{eq:corrx}
\ee
This behavior is readily observed in Fig.~\ref{fig:corr} where panels (a) and (b) show a comparison between average and typical decays for a few representative values of the interaction strength. The extracted exponent governing the average is fully consistent with the universal IRFP value $\eta=2-\phi$~\cite{fisher_random_1992}, where $\phi$ is the golden mean. The typical decay, while suffering from finite size effects, also appears to be in good agreement with a stretched exponential vanishing.

%%%%%%%
%%%%%%%
\section{Incommensurability} 
\begin{figure}[bp]
    \centering
    \includegraphics[width=.88\columnwidth]{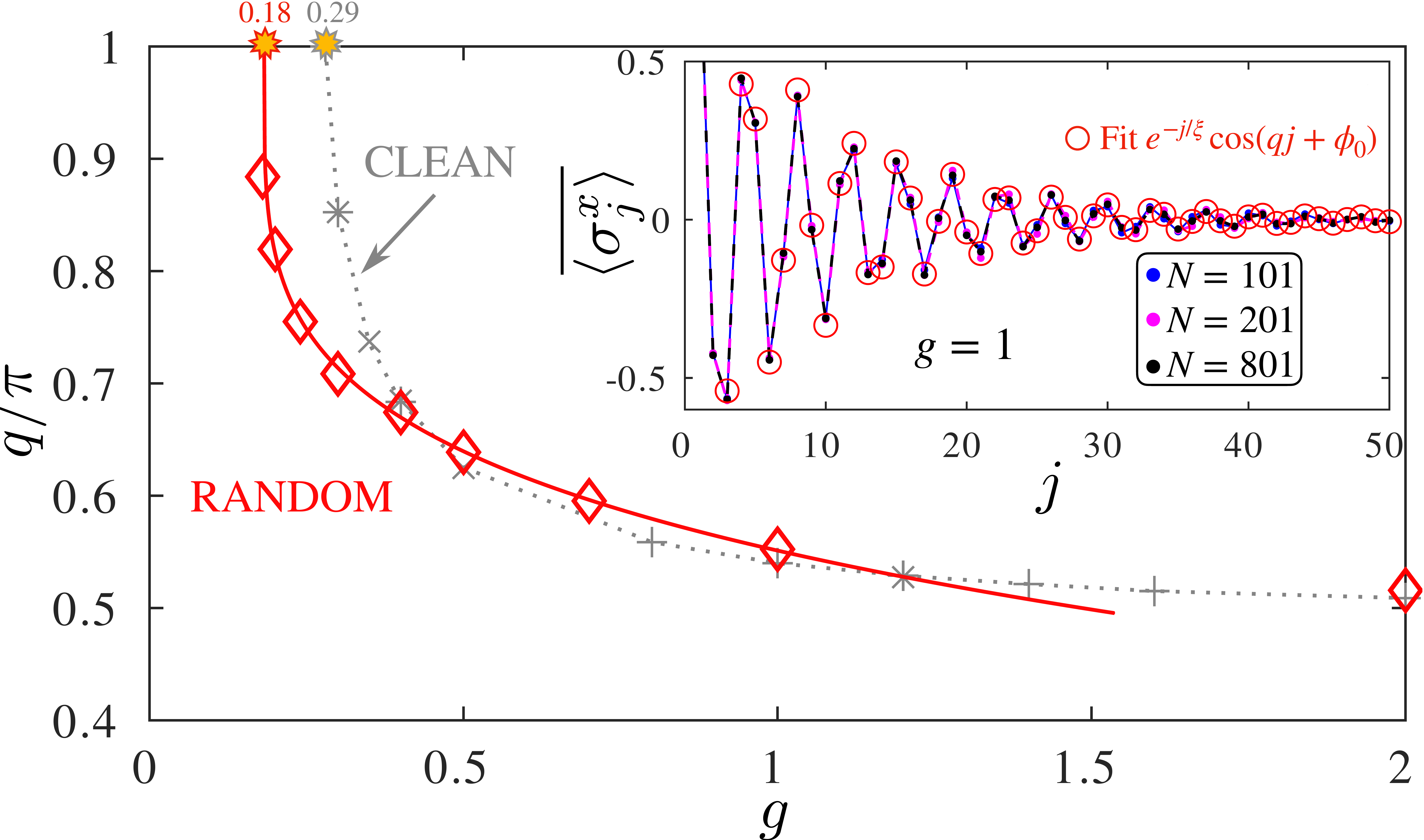}
        \caption{IC wave-vector $q$ extracted from exponential scaling of average magnetization away from the polarized boundary ($g^\star\approx 0.18$, red symbols), compared to the clean case~\cite{chepiga_topological_2022} ($g^{\star}_{\rm clean}\approx 0.29$, grey symbols).
        Inset: Average magnetization decay away from the boundary shown at $g=1$ for various lengths $N$ and compared to the fit.} \label{fig:IC}
\end{figure}

So far we have focused on the absolute value of the magnetization, 
ignoring possible commensurate or incommensurate (IC) modulations. However, while the mean of the absolute value ${\overline{|\langle \sigma_{j}^{x}\rangle|}}$ does decay algebraically, the mean magnetization vanishes much faster ${\overline{\langle \sigma_{j}^{x}\rangle}}\propto \exp\left(-j/\xi\right)\cos(qj)$, with antiferromagnetic correlations ($q=\pi$) for $g\le g^\star$, which then turns IC ($\pi/2<q<\pi$) beyond $g^\star\approx 0.18$, see  Fig.~\ref{fig:IC}.
It is noteworthy that the IC behavior induced by the frustrating nature of the interaction is not pinned by the disorder, as previously suggested~\cite{milsted_statistical_2015}, but actually seems to be enhanced compared to the clean case for which $g^{\star}_{\rm clean}\approx0.29$~\cite{chepiga_topological_2022}.
Nevertheless, IC is only short-range because the Luttinger liquid is localized by the disorder.
%%%%%%%

\section{Discussions and conclusions} In the strong-disorder RG (SDRG) framework~\cite{fisher_random_1992,fisher_critical_1995,igloi_strong_2005}, adding (moderate) interactions to the random-bond XX chain only brings negligible modifications to the RG recursion relations, and the IRFP has the very same form as in the non-interacting XX case, notably for the Heisenberg chain~\cite{fisher_random_1994}. However, this is less obvious for the interacting version of the TFIM, as recently discussed by Monthus~\cite{monthus_strong_2018} who showed that the SDRG treatment of  disordered interacting Majorana fermions generates higher-order couplings, which prevents direct conclusions about the effects of interactions, a situation also encountered for more general random XYZ models~\cite{roberts_infinite_2021} as well as for MBL~\cite{pekker_hilbert-glass_2014,slagle_disordered_2016,you_entanglement_2016}.

In such a puzzling context, our numerical work substantially clarifies the problem, providing a simple picture which contrast with previous works~\cite{milsted_statistical_2015,karcher_disorder_2019}. Building on state-of-the-art DMRG simulations, appropriate boundary conditions, and a very large number of samples, we demonstrate that the non-interacting  IRFP is stable against attractive and repulsive interactions between Majorana fermions. This solves a relatively old problem, and open interesting questions regarding the  stability of the marginally localized~\cite{nandkishore_marginal_2014} IRFP far from the ground-state where instead, weak interactions are expected to delocalize and restore ergodicity,  at least in the infinite-temperature limit~\cite{sahay_emergent_2021,moudgalya_perturbative_2020,laflorencie_topological_2022}, thus suggesting a possible critical point at finite energy density above the ground-state. 

\vskip 0.15cm
\section{Acknowledgments}
We thank J. Hoyos and I. C. Fulga for comments. NC acknowledges LPT Toulouse (CNRS) for hospitality. This work has been supported by Delft Technology Fellowship, by the EUR grant NanoX No. ANR-17-EURE-0009 in the framework of the ''Programme des Investissements d'Avenir''.
Numerical simulations have been performed at the DelftBlue High Performance Computing Centre (\href{https://www.tudelft.nl/dhpc}{tudelft.nl/dhpc}) and CALMIP (grants 2022-P0677). 
%%%%%%%%%%%%%%%

\vskip 1cm
\setcounter{secnumdepth}{3}
\setcounter{figure}{0}
\setcounter{equation}{0}
\renewcommand\thefigure{S\arabic{figure}}
\renewcommand\theequation{S\arabic{equation}}

\begin{center}
    \bfseries\large Supplemental material
    % to ``Resilient infinite randomness criticality for a disordered chain of interacting  Majorana fermions''
\end{center}

\vskip 0cm

\section{Models and useful transformations}
\begin{figure}[h!]
\centering 
\includegraphics[width=.95\columnwidth]{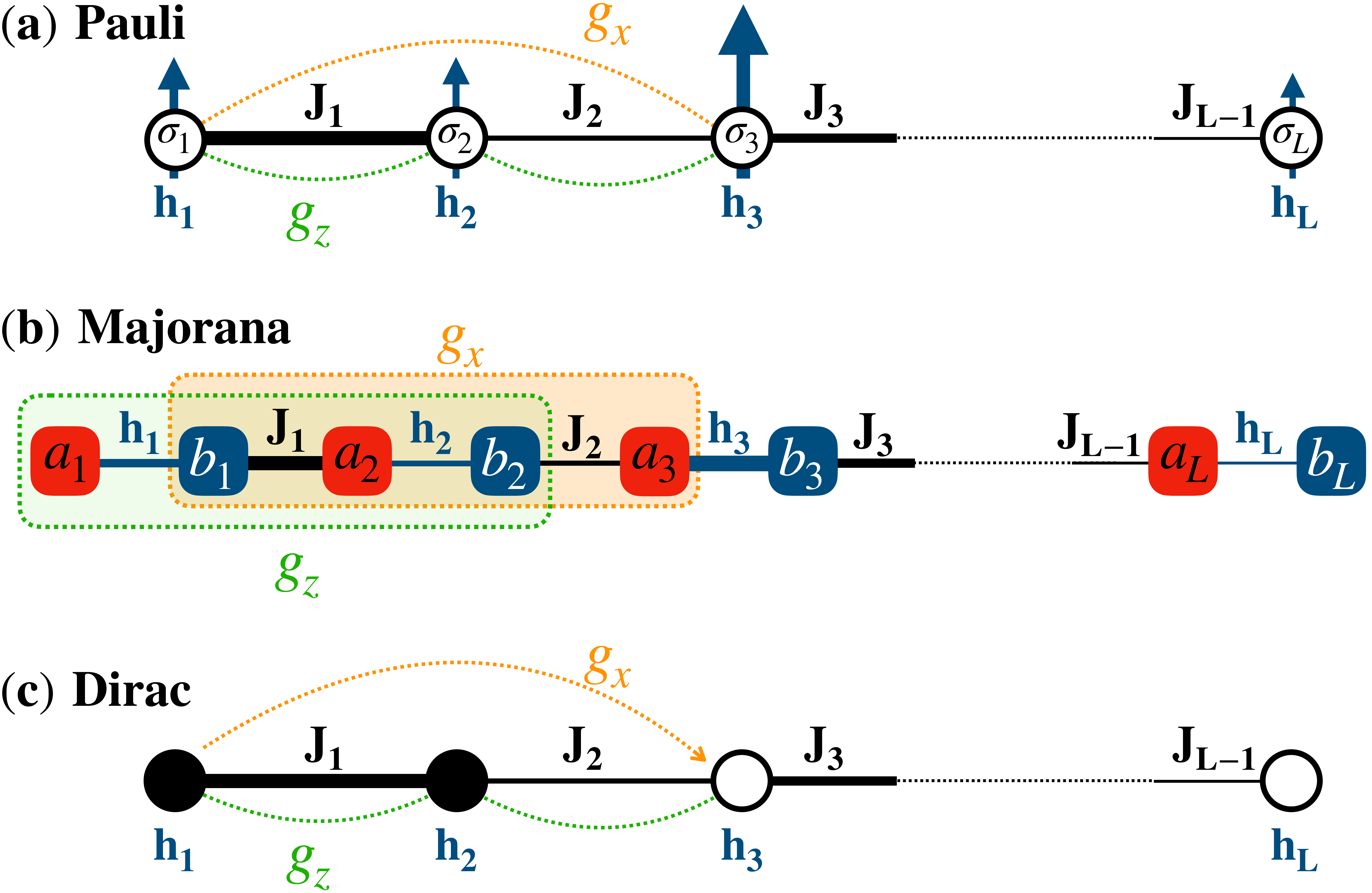}
\caption{Schematic picture for the same interacting Majorana chain model. (a) Spin  Hamiltonian Eq.~\eqref{eq:spin}, (b) Majorana fermion representation Eq.~\eqref{eq:majorana}, (c) Dirac fermions Eq.~\eqref{eq:KitaevI}.}
\label{fig:sketch}
\end{figure}

The interacting Majorana chain model studied in the main text  is governed by the one-dimensional Hamiltonian
\be
{\cal{H}}=-\sum_j\left({\rm{i}}t_j\gamma_{j}\gamma_{j+1}+g\gamma_{j}\gamma_{j+1}\gamma_{j+2}\gamma_{j+3}\right),
\label{eq:s1}
\ee
with random couplings $t_j$ and constant interaction $g$. It is more convenient to introduce odd and even Majorana operators 
$\gamma_{2j-1}=a_\ell\quad{\rm{and}}\quad \gamma_{2j}=b_\ell$, 
%which satisfy the usual rules $(a/b)^{\dagger}=(a/b)$, $(a/b)^2=1$, $\{a_\ell,a_{\ell'}\}=\{b_{\ell},b_{\ell'}\}=2\delta_{\ell\ell'}$, and $\{a_\ell,b_\ell\}=0$).
These new operators are  connected to the real space lattice sites $\ell$ where live  Dirac fermions and Pauli matrices operators. We use the Jordan-Wigner mapping
\bea
a_\ell&=&c_\ell^\dagger+c_\ell^{\vphantom{\dagger}}=K_\ell\sigma^x_\ell\\
b_\ell&=&{\ii}(c_\ell^\dagger-c_\ell^{\vphantom{\dagger}})=K_\ell\sigma^y_\ell\\
a_\ell b_\ell &=&{\ii}(1-2c_\ell^\dagger c_\ell^{\vphantom{\dagger}})={\ii}\sigma_\ell^z\\
{\rm{with}}\quad K_{\ell}&=&\prod_{k=1}^{\ell-1}\sigma_k^z,
\eea
such that the above interacting Majorana chain can be expressed in three languages: Pauli, Majorana, and Dirac, as sketched in Fig.~\ref{fig:sketch}. It is also instructive to inroduce the possibility for asymmetric interactions $g_{x,z}$,  such that Eq.~\eqref{eq:s1} reads
\bea
{\cal{H}}_{\rm Majorana}&=&-{\ii}\sum_\ell \left(J_\ell b_\ell a_{\ell+1}-h_\ell a_\ell  b_\ell \right)\\
&-&\sum_\ell \left( g_z a_\ell  b_\ell  a_{\ell+1}b_{\ell+1}+g_x  b_\ell  a_{\ell+1}b_{\ell+1}a_{\ell+2}\right),\nonumber
\label{eq:majorana}
\eea
with $J_{\ell}=t_{2j}$ and $h_{\ell}=t_{2j-1}$, as sketched in Fig.~\ref{fig:sketch} (b). In the Pauli (spin) language, the same model becomes
\bea
{\cal{H}}_{\rm Pauli}&=&\sum_\ell\left(J_\ell\sigma^x_\ell\sigma^x_{\ell+1}-h_\ell\sigma^z_\ell\right)\nonumber\\
&+&\sum_{\ell}\left(g_z \sigma^z_\ell\sigma^z_{\ell+1}+g_x \sigma^x_\ell\sigma^x_{\ell+2}\right)
\label{eq:spin}
\eea
where $\sigma^{x,z}_\ell$ are Pauli matrices at site $\ell$, see Fig.~\ref{fig:sketch} (a). Finally, in terms of Dirac fermions, we have the interacting version of the Kitaev chain model~\cite{kitaev_unpaired_2001}, illustrated in Fig.~\ref{fig:sketch} (c)
\bea
{\cal{H}}_{\rm Dirac}&=&\sum_\ell\Bigl[J_{\ell}\left(c_\ell^{\dagger}c_{\ell+1}^{\vphantom{\dagger}} + c_\ell^{\dagger}c_{\ell+1}^{{\dagger}} + {\rm{h.c.}}\right)+2h_\ell n_\ell\Bigr]\nonumber\\
&+&g_z\sum_\ell\left(1-2n_\ell\right)\left(1-2n_{\ell+1}\right)\\
&+&g_x\sum_\ell\left(c_\ell^{\dagger}-c_{\ell}^{\vphantom{\dagger}}\right)\left(1-2n_{\ell+1}\right)\left(c_{\ell+2}^{\dagger}+c_{\ell+2}^{\vphantom{\dagger}}\right).\nonumber
\label{eq:KitaevI}
\eea
The $g_z$ coupling is a simple density-density interaction term at distance 1: $\sim g_z n_\ell n_{\ell+1}$. Instead, the $g_x$ coupling brings frustration to the problem and displays the very interesting density-assisted hopping $\sim g_x c_{\ell}^{\vphantom{\dagger}} n_{\ell+1} c_{\ell+2}^{\dagger}$ and pairing $\sim g_x c_{\ell}^{{\dagger}} n_{\ell+1} c_{\ell+2}^{\dagger}$ terms at distance 2.
%%%%%%%%%%%%%%%%%%%%%
%%%%%%%%%%%%%%%%%%%%%
%

\section{Entanglement entropy distribution}
\begin{figure}[h!]
    \centering
    \includegraphics[width=.95\columnwidth]{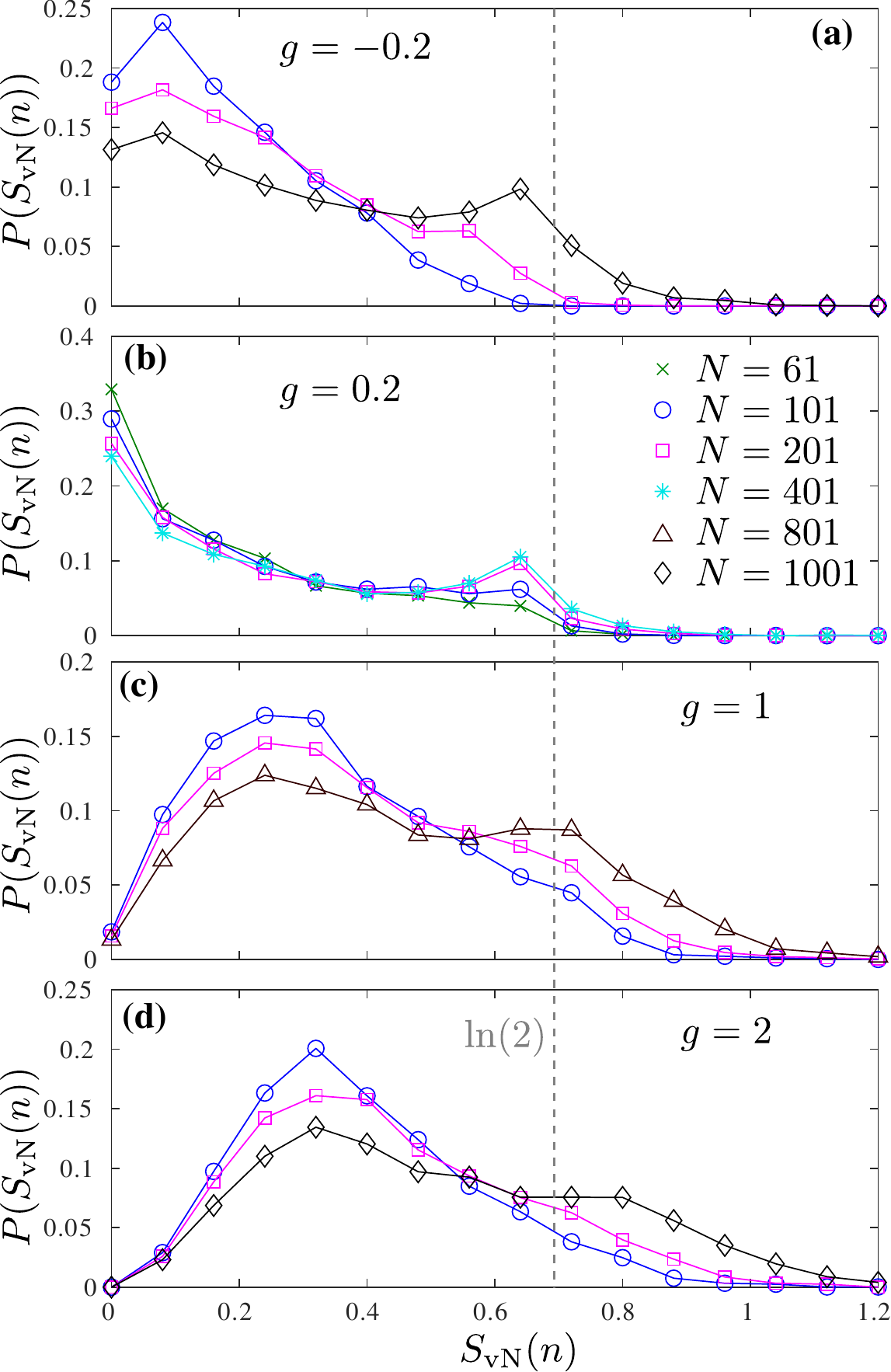}
        \caption{Entanglement entropy distribution for four different coupling contants $g$ and various system sizes. The probability develops a peak at $\ln(2)$ typical for infinite randomness critical point. The peak at zero for weak interaction is gradually depleted with increasing $N$ (a-b), while it is already quasi-absent at large $g$ (c-d) because the disorder-free interaction terms prevent to have cut with $S_{\rm vN}=0$.} \label{fig:EE_hist}
\end{figure}

Here we show several examples of  middle-chain entanglement entropy distributions in Fig.~\ref{fig:EE_hist} for various interaction strengths and system sizes. Upon increasing $N$, one sees a slow crossover towards IRFP physics signalled by a peak at $S_{\rm vN}\approx \ln 2$~\cite{refael_entanglement_2004,laflorencie_scaling_2005}. Interestingly, for $g=-0.2$ and $g=0.2$ there is also a peak at zero entropy, but it  slowly decreases with growing $N$ while  its weight is  transferred to $\ln 2$. 
This is not observed for $g = 1,\, 2$ due to this large coupling strength, which prevents zero entanglement (see e.g. Fig.~\ref{fig:sketch} (a) with $g_x = g_z$).

%%%%%%%%%%
\vskip 0.25cm

\section{Additional results on incommensurate correlations}
\begin{figure}[h!]
    \centering
    \includegraphics[width=.87\columnwidth]{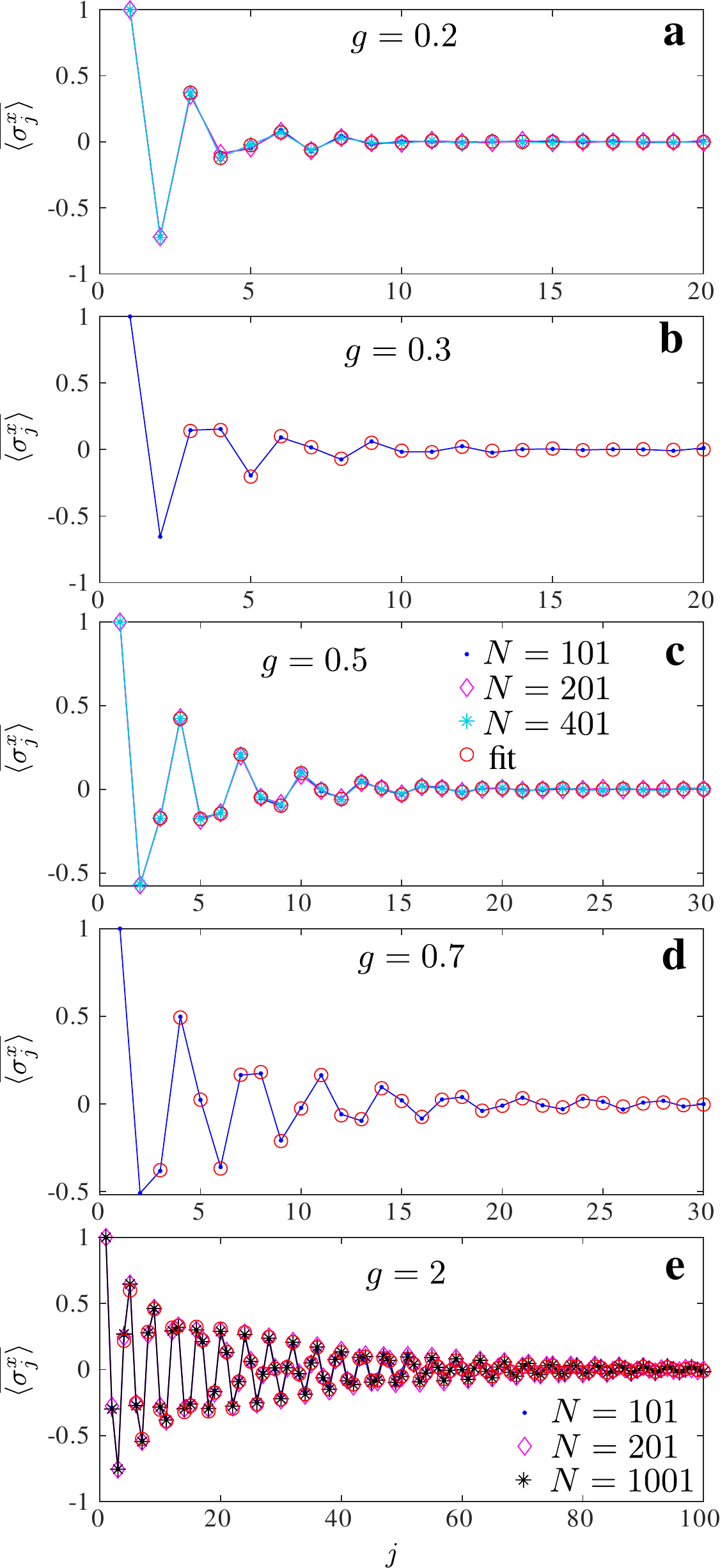}
        \caption{Disorder-average magnetization in the vicinity of the polarized edge spin for $N=101$ (blue), $N=201$ (magenta), $N=401$ (cyan) and $N=1001$ (black) for various values of the coupling contant $g$.  Red open circles are fit with $\propto e^{-j/\xi}\cos(q\cdot j+\phi_0)$. The extracted values of $q$ are summarized in Fig.~\ref{fig:IC} of the main text.} \label{fig:IC_sm}
\end{figure}

In Fig.\ref{fig:IC_sm} we provide additional numerical results for incommensurate Friedel oscillations that appear as a response to  a boundary spin  polarized in $x$-direction. We show results for five different values of the coupling constant  ranging from $g=0.2$ that in the clean case is located below the Lifshitz point (i.e. in the region where the correlations are still commensurate) to the strongly-interacting case  $g=2$, where the extracted wave-vector $q\approx0.515\pi$  fully agrees (the difference is less than $1\%$) with the value of $q$ in the clean case.

%%%%%%%%%%
%%%%%%%%%%

\newpage\section{Extracting energy gaps with DMRG}
In order to extract the energy gap, we target several low-lying eigenstates of the effective Hamiltonian at every DMRG iteration and keep track of the energy as a function of iteration. Following Ref.~\cite{dmrg_chepiga} we associate reliable energy levels with energies that remains flat for several DMRG iterations. In practice, the flat intervals span over almost the entire chain length except very close to the edges, where the effective basis is known to be too small to properly capture an excited state. Such an excellent convergence of the excitation energy for disordered systems is quite surprising but systematically good from sample to sample. Probably the reason behind it is an infinite correlation length, for this the used method is known to be extremely accurate\cite{dmrg_chepiga,10.21468/SciPostPhysCore.5.2.031}.
 The approach is troublesome for very small gaps (of the order of $10^{-8}$ and below), however considered values of $g$ this has a noticeable contribution on a disordered chains with length above $80-100$ sites.

\begin{figure}[h!]
    \centering
    \includegraphics[width=\columnwidth]{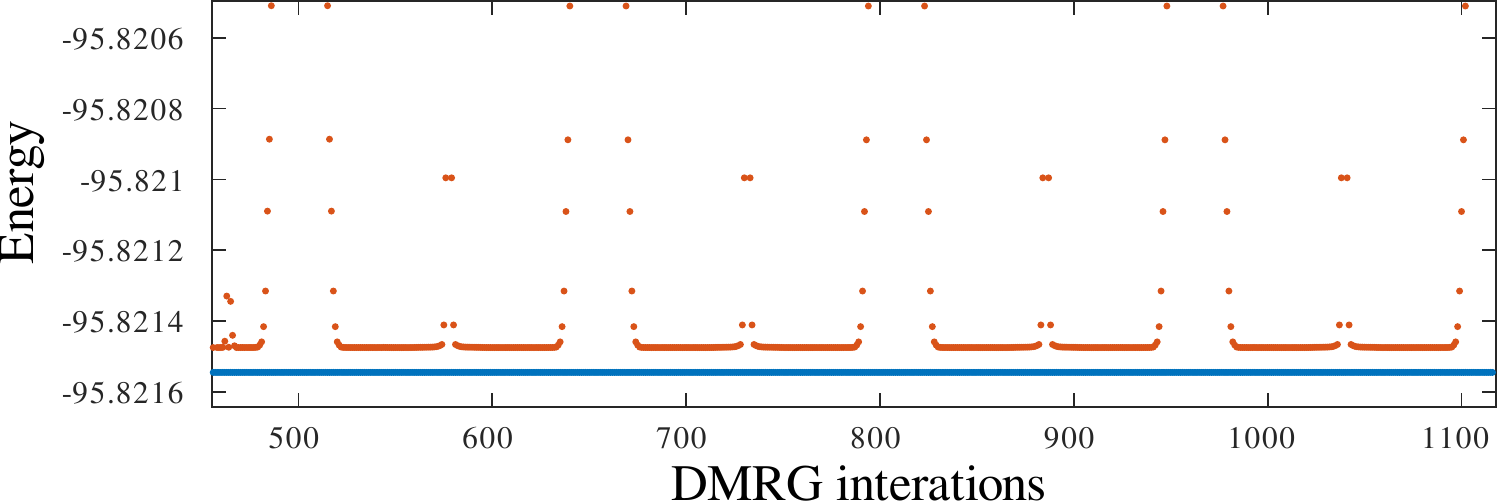}
        \caption{Example of a DMRG convergence of the two lowest energy levels for a selected sample with $N=80$, $g=0.5$ as a function of DMRG iterations. Regions where the first excitation energy (red dots) is flat as a function of DMRG iterations corresponds to a reliable value of the excitation energy. Periodic increase of the energy of the first excited state takes place close to the edges of the chain and is a typical artifact of the method. In the presented case the energy gap is $\Delta\approx 2\cdot 10^{-5}$.} \label{fig:cvg_sm}
\end{figure}
%%%%%%%%%%%%%%
\bibliography{biblio,notes}

\begin{thebibliography}{57}
\bibitem [{note_dmrg()}]{note_dmrg}%
  \BibitemOpen
  \bibinfo {note} {In our DMRG simulations we keep up to $D=3000$ states,
  perform from 4 to 8 sweeps and discard singular values lower than $10^{-8}$.
  In order to avoid convergence to a stable local minima at the low
  interactions ($g\leq 0.2$) we target multiple states in the Lanczos
  subroutine that prevents the system from being stacked at the low-lying
  excited state.}\BibitemShut {Stop}%
\end{thebibliography}

\end{document}